\RequirePackage[hyphens]{url}

\documentclass[%
reprint,
superscriptaddress,
 amsmath,amssymb,
 aps,
 prab,
floatfix,
]{revtex4-2}

\usepackage{graphicx}
\usepackage{dcolumn}
\usepackage{bm,booktabs}
\usepackage{amsthm}
\usepackage{multirow}
\usepackage{color, soul}
\usepackage{hyperref}
\usepackage{tikz}
\usetikzlibrary{shapes.geometric, arrows}

\theoremstyle{definition}

\tikzstyle{io} = [rectangle, minimum width=3cm, minimum height=1cm, text centered, text width=3cm, draw=black]
\tikzstyle{nn} = [rectangle, minimum width=3cm, minimum height=0.5cm, text centered, text width=6cm, draw=black]
\tikzstyle{arrow} = [thick,->,>=stealth]

\newcommand{\AffilICME}{\affiliation{%
 Institute for Computational and Mathematical Engineering, Stanford University\\
 Stanford, California 94305, USA
}}
\newcommand{\AffilSLAC}{\affiliation{%
 SLAC National Laboratory\\
 Menlo Park, California 94025, USA
}}

\begin{document}

\preprint{APS/123-QED}

\title{Beam-based RF Station Fault Identification at the SLAC Linac Coherent Light Source}

\author{Ryan Humble}%
 \email{ryhumble@stanford.edu}%
 \AffilICME
\author{Finn H. O'Shea}%
 \AffilSLAC
\author{William Colocho}%
\author{Matt Gibbs}%
\author{Helen Chaffee}%
 \AffilSLAC
\author{Eric Darve}%
 \altaffiliation[Also at ]{Department of Mechanical Engineering, Stanford University.}%
 \AffilICME
\author{Daniel Ratner}%
 \AffilSLAC

\date{\today}

\begin{abstract}
Accelerators produce too many signals for a small operations team to monitor in real time.  In addition, many of these signals are only interpretable by subject matter experts with years of experience.  As a result, changes in accelerator performance can require time-intensive consultations with experts to identify the underlying problem.  Herein, we focus on a particular anomaly detection task for radio-frequency (RF) stations at the SLAC Linac Coherent Light Source (LCLS). The existing RF station diagnostics are bandwidth limited, resulting in slow, unreliable signals. As a result, anomaly detection is currently a manual process. We propose a beam-based method, identifying changes in the accelerator status using shot-to-shot data from the beam position monitoring system; by comparing the beam-based anomalies to data from RF stations, we identify the source of the change.  We find that our proposed method can be fully automated while identifying more events with fewer false positives than the RF station diagnostics alone. Our automated fault identification system has been used to create a new data set for investigating the interaction between the RF stations and accelerator performance.

\end{abstract}

\maketitle


\section{\label{sec:intro}Introduction}
The increasing complexity of particle accelerators is creating new challenges in operating user facilities with high-uptime. The tasks of predicting, identifying, and recovering from anomalies has already grown beyond what a human operator can monitor unassisted ~\cite{fazio_accelerator_needs}; for example, the control system for SLAC's Linac Coherent Light Source (LCLS) includes 200,000 different process variables (PV) streams, while the next-generation LCLS-II will record 2 million. This growth in data rates opens new opportunities for data science techniques to improve automation of accelerator operations.

The problem of anomaly detection, the task of finding abnormal events or data, is common beyond particle accelerators. Nearly all modern data-heavy systems, such as complex internet, industrial, and engineering systems, produce thousands of real-time signals, overwhelming the human operators tasked with monitoring performance. Moreover, research has shown that humans, even trained operators, are severely limited when detecting rare events~\cite{drew_inattentional_blindness, simons_inattentional_blindness, wolfe_rare_missed}. Nonetheless, human operators have the necessary causal reasoning and deep institutional knowledge to investigate and understand the root cause of anomalies. To better detect anomalies in these complex systems, we seek to delegate the constant monitoring of a large number of complex data streams to an automated AI-agent and refocus human attention, cognition, and causal reasoning to anomaly investigation and recovery.

\subsection{\label{sec:lcls_intro}Anomalies at LCLS}
LCLS, like many complex engineering systems, experiences unplanned downtime and unexpected behavior that is difficult to pinpoint and correct. LCLS loses at least $3\%$ of availability\footnote{Only downtime events lasting longer than $ 6 $ minutes are tracked. The downtime statistics are therefore underestimates.} -- more than $180$ hours per year -- to unplanned downtime, equivalent to more than three full user experiments. We also focus on more subtle changes to performance which degrade the free-electron laser (FEL) performance, even when the beam is still delivered.  Furthermore, fluctuations in performance can restrict the operational range of the accelerator (e.g., if spare RF stations must be held in reserve, the range of X-ray energy available to users is limited) or contribute noise to experiments through changes in X-ray photon energy. Therefore, our goal is to minimize the number of interruptions to stable beam, rather than integrated downtime. Due to limited bandwidth across the multi-kilometer accelerator, subsystem diagnostics are typically low-fidelity, and existing methods for identifying faults purely on subsystem diagnostics have both too many false positives and false negatives to be deployed in operation.

The motivation for this work is three-fold: (i) develop an algorithm that will alert an operator that an anomaly has occurred and propose a candidate source; (ii) automate logging of anomalous events; and (iii) create a new labeled anomaly data set that can be used for future studies.

First, by alerting operators that an anomaly has occurred and localizing a candidate source, we aim to increase up-time and reliability, decrease time to restore the beam to user operation, lower maintenance costs, and, indirectly, expand the effective operational limits of the machine. Second, automating the anomaly detection process enables automated logging of events, which can in turn identify sub-systems in need of repair prior to failures or sub-systems that are recurrently problematic.
These goals align with ongoing work at other organizations, such as the Smart LINAC project at CERN \cite{smart_linac}, which is utilizing AI-based anomaly detection to maximize up-time and to decrease maintenance costs for LINACs. 

Finally, the creation of labeled, voluminous, data sets is often the first step in tackling any machine learning (ML) problem~\cite{Domingos:2012uuu}.  Thus, in addition to the operational benefits just mentioned, an algorithm that can automatically detect anomalous behavior can also enable the creation of a labeled data set. The problem of assembling data sets has been mentioned in a number of recent works on applying ML at accelerators.  Specifically, they point to burden created by the amount of effort required to produce a suitable data set~\cite{Donon:2019uuu,Tennant2020,Wielgosz:2017uuu}, or they expressed a desire for more data to improve the performance of their algorithms~\cite{Li2021,Rescic2020}. In the RF station example, a dataset labeled with slow RF station data could enable a future algorithm based entirely on fast, beam-based signals. 

\subsection{Contribution}
In this work, we introduce a fully automated, beam-based diagnostic method for RF station fault detection at LCLS. Our key contribution is the combination of high fidelity beam position monitoring data with bandwidth limited (and noisy) RF station diagnostic data to achieve highly accurate, nearly real time fault identification. This approach, of using beam-based data as a way of understanding subsystem performance, is applicable beyond RF stations at LCLS to anomaly problems at LCLS-II and other accelerators. We primarily present an interpretable unsupervised anomaly detection algorithm that exploits knowledge of LCLS's design and seeks to mimic a human operator's fault analysis process.

Our paper's contributions are: 
\begin{enumerate}
    \item We describe the current problem of identifying RF station faults at LCLS, notably the bandwidth limitation and inaccuracy of the existing RF station diagnostics system, and explain how pulse-by-pulse beam position data can be used to supplement these diagnostics. See Sec.~\ref{sec:klystron_faults}.
    \item We present how to identify possible RF station faults, which we refer to as anomaly candidates, from two primary existing RF station diagnostics. See Sec.~\ref{sec:cand_gen}.
    \item We define a statistics-based unsupervised time-series anomaly detection algorithm on the beam position data. The algorithm assigns an anomaly score to each anomaly candidate to determine if a RF station fault occurred. See Sec.~\ref{sec:fault_confirm}.
    \item We compare our fully automated system to an existing detection system, over a labeled study period, and find our system reduces the number of false positives by a factor of \(20\)x while correctly identifying \( 96\% \) of the faults. Over a longer unlabeled period, we analyze the faults identified by our approach and correlate our findings with faults that were manually logged by LCLS operators. Lastly, we describe the creation of a public anomaly detection dataset from the RF station diagnostics, beam position data, and hand labels, and demonstrate that a proof of concept DNN trained on this data achieves comparable results to our original (unsupervised) system. See Sec.~\ref{sec:results}.
\end{enumerate}


\section{Related work}




Fault identification and recovery reduces downtime and improves performance metrics.  Within the last 5 years, various accelerators have investigated machine learning algorithms to identify unusual (i.e. anomalous) accelerator states to increase the performance of the accelerator: isolation forests to improve the quality of optics measurements by detecting faulty BPMs \cite{Fol2020}; autoencoders to show sensitivity to parameter drifts in drift-tube LINACs \cite{Edelen2021}; Gaussian mixture models and isolation forests to detect anomalous behavior in injection kicker magnets \cite{Dewitte:2019uuu}; transfer learning to detect anomalous vacuum pressure changes at a synchrotron \cite{Piekarski2020}; principal component analysis and local outlier factor to detect collimation system settings that need improvement \cite{Valentino2017}; Siamese neural network model to predict anomalous pulses at a proton LINAC and prevent damage to the accelerator \cite{Blokland2021}; and recurrent autoencoders to detect failing RF power modulators \cite{Radaideh2021}.  \cite{Tilaro2017} applies a variety of different model-based and non-model-based techniques for detecting anomalies in a cryogenic system, including an online update to the definition of anomaly.  \cite{Donon:2019uuu,Donon:2020uuu} also apply a number of methods to detect anomalies of different types in the RF power of a proton source.  Several algorithms have been used to monitor the cryogenic system at the European XFEL, such as a residual of the eigenvalues of a state matrix to detect quenches in superconducting cavities at reduced repetition rate \cite{Nawaz:2016uuu}, and anomaly detection using a model-based parity space algorithm and classification using support vector machines \cite{Nawaz:2018xyz,Nawaz:2018qrs}.  In the latter case, anomalies are defined on an ad-hoc basis by separating clusters.



\section{\label{sec:klystron_faults}RF Station Anomalies}
The health of LCLS is judged by the quality of the X-ray beam delivered to users, typically through metrics such as energy per pulse, stability of photon energy, and X-ray bandwidth. For the purposes of this work, we define an anomaly to have occurred when there is an unexpected change in the state of the accelerator that either degrades the performance of the X-ray beam or causes a machine fault that halts operation entirely. After observing the onset of degraded performance, operators re-tune based on previous operational experience, a collection of physical models, and intuition.  With strong anomalies the machine-protection system will turn off beam delivery, and the same techniques are used to recover accelerator operation. Searching for the root-cause of an anomaly can be time-consuming, both due to the large number of process variables and due to the complex interaction between the accelerator and its various automated control systems, which may blur cause and effect.


Out of LCLS' various failure modes, we focus on the challenge of identifying anomalous behavior in the $82$ RF stations that power the X-ray laser’s accelerator. Changes in the power delivered by an RF station can range from nearly imperceptible impact to the electron beam to completely stopping operation (i.e., a machine fault). Fault identification and recovery currently requires an experienced operator searching through hundreds of data streams in real time after the fault occurs. Anomalies in the RF stations are an interesting source of study for multiple reasons: (i) they are a major failure mode, with several drops in RF station performance per hour during beam operation; (ii) they directly affect the beam energy, causing degradation of free-electron laser (FEL) performance; and (iii) they have an existing but inaccurate detection system.





\subsection{\label{sec:existing_method}Current Method}
There are two existing methods for detecting RF station faults at LCLS. A drop in RF station performance may be observed either through (i) system diagnostics that directly monitor the RF station, or (ii) through the subsequent impact on the accelerator itself.

Unfortunately, the RF station diagnostics have two shortcomings: the data is reported no faster than once every 5 seconds (600x slower than the beam rate) due to limited bandwidth in the control system; and, more fundamentally, the diagnostic info is often incorrect, falsely reporting normal performance as anomalous and missing real dips in performance (as shown in Section~\ref{sec:fault_ident_result}).

Consequently, LCLS primarily uses a human-centered approach following a drop in accelerator performance, with operators checking by hand a set of RF station diagnostics for a possible cause of the anomalous behavior. This human-driven fault detection and recovery has drawbacks, requiring the attention of a human operator for a routine activity. Moreover, because the search for an anomaly is prompted by an operator, the human-centered approach only addresses significant and prolonged changes in performance that catch the operator's attention, and misses less-noticeable -- but possibly frequent -- drops in performance that may still impact user experiments or may predict future failures.

\subsection{\label{sec:lcls_data_sources}Combining LCLS Data Sources}
As an alternative to the human-centered approach, we combine two sources of data available at LCLS: RF station diagnostic data, which are sensitive to the source of the fault, and beam-based diagnostics, which are sensitive to pulse-by-pulse accelerator performance. Combining the two data streams makes it possible to, accurately and with fully autonomy, both detect of RF station anomalies and identify the responsible RF station, thereby avoiding the shortcomings of the inaccurate and manual current method.

The RF station diagnostic data consists of roughly a dozen status bits (0 - healthy, 1 - unhealthy) as well as several real-valued signals. Most notably, there is a klystron AMM (amplitude mean out of tolerance) status bit that reports if the station klystron's accelerating RF amplitude is too high or low and a downsampled version of the real-valued AMPL (amplitude) signal. Klystron amplitude is related to the energy gain of the electron bunch so a change in amplitude will change the energy of the bunch and degrade the ability to transport the bunch through the magnetic lattice. The amplitude is derived from directional couplers at the rectangular waveguides that feed the accelerating structures. When searching for a fault by hand, operators typically use the AMM status bit. However, since both the AMM and AMPL are inaccurate, we cannot rely on this source of data alone. We continue to explore using the unused RF station diagnostic information, including phase information PHAS, for improved anomaly detection.

The second source is from beam-based diagnostic data.
The human-centered evaluation of FEL performance leans on X-ray diagnostics, which measure the quantities of interest for the FEL users. However, X-ray performance is sensitive to nearly every potential type of failure.  Instead, we focus on stripline beam position monitor (BPM) \cite{Strausmann:2007uuu} data in dispersive regions, which are sensitive primarily to fluctuations in beam energy. The BPM data is included in the beam-synchronous acquisition service (BSA)~\cite{kim_bsa}, which aligns a subset of the machine signals to the individual beam pulses; this enables BPM data to be read and recorded at the full LCLS operational rate of 120~Hz. As the ``ground truth'' for the beam status, the BPM data allows anomalies to be discovered at the full beam rate; however, as the BSA service does not record any RF station diagnostic information, attributing the anomaly to a particular component of the accelerator is challenging.

\begin{figure*}[htbp]
\centering
\includegraphics[width=\linewidth]{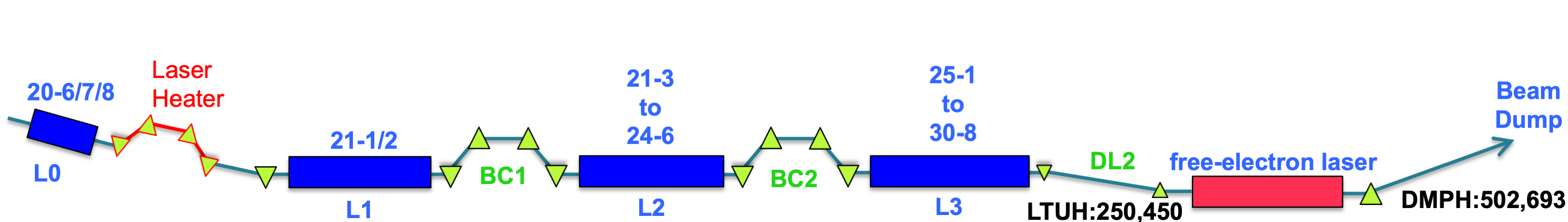}
\caption{\label{fig:lcls_layout} Layout of the LCLS showing the location of the klystrons included in this work above the accelerating sections (blue rectangles: L0, L1, L2, and L3) and the names of the BPMs used to detect changes in beam energy: LTUH:250, LTUH:450, DMPH:502, and DMPH:693. The green triangles show dipoles and the red rectangles represent the free-electron laser undulators.}
\end{figure*}

Each of the 175 BPMs along the accelerator produces three signals: X, Y, and TMIT. X and Y measure transverse position; TMIT (transmitted intensity) measures the passing beam charge. Of the 175 BPMs, there are 4 BPMs of interest to our work: BPMS:LTUH:250, BPMS:LTUH:450, BPMS:DMPH:502, and BPMS:DMPH:693. As shown in Figure~\ref{fig:lcls_layout}, all of these BPMs are located in regions of the accelerator with non-zero dispersion in either X or Y, making them sensitive to changes in beam energy, and are downstream of all RF stations. We take the 4 beam positions in the dispersive plane and the 4 TMIT signals as our input BPM data. 

We then seek to use the BPM data to corroborate and enrich the RF station status that is reported by the diagnostic data. By cross-referencing the faster, more sensitive BPM data with the slower, easier-to-interpret diagnostic data, the algorithm adopts the strengths of both: changes in the RF station diagnostic that are not corroborated by the BPMs are rejected, and subtle changes to the RF station that were previously rejected can be confirmed by the BPMs.  We evaluate the performance of this strategy on archival data at LCLS.

\section{\label{sec:method}Method}
\begin{figure}[ht]
    \centering
    \begin{tikzpicture}[node distance=2cm]
        \node (rfs_in) [] {RF station diagnostic data};
        \node (cand_ident) [io, below of=rfs_in] {Anomaly candidate identification\\(Section~\ref{sec:cand_gen})};
        \node (anomaly_confirm) [io, below of=cand_ident] {Anomaly\\confirmation\\(Section~\ref{sec:fault_confirm})};
        \node (bsa_in) [left of=anomaly_confirm, xshift=-2cm] {Beam data};
        \node (candidates) [below of=anomaly_confirm] {Confirmed anomalies};
    
        \draw [arrow] (rfs_in) -- (cand_ident);
        \draw [arrow] (cand_ident) -- (anomaly_confirm);
        \draw [arrow] (bsa_in) -- (anomaly_confirm);
        \draw [arrow] (anomaly_confirm) -- (candidates);
    \end{tikzpicture}
    \caption{Two-stage RF station anomaly identification method.}
    \label{fig:method_diagram}
\end{figure}
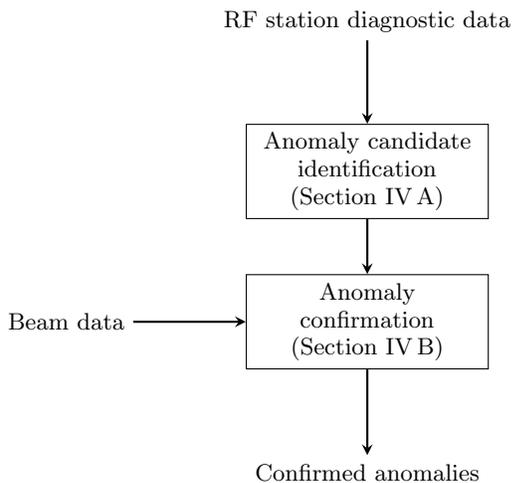

Our method can be broken down into two steps. The first step is anomaly candidate identification, where we use the slowly-updating RF station diagnostic data to identify possible anomalies. The second is to confirm the anomaly using the full-rate BPM data. A flow chart for this process is shown in Figure~\ref{fig:method_diagram}.



\subsection{\label{sec:cand_gen}Anomaly Candidate Identification}
The first step generates candidates independently for each RF station based on each station's diagnostic data.  Formally, given diagnostic data $ h_{k,t} $ from RF station $ k $ at time $ t $, we want to identify anomaly candidates $ C = \{ C_i \} $. A candidate is \( C_i = (s_i, e_i, k_i) \) for start time $ s_i $, end time $ e_i $, and station $ k_i $. Candidates are merged across all RF stations with overlapping anomaly windows. We then drop candidates associated with more than one RF station for the sake of interpretability of the performance metrics.  This yields a set of anomaly candidates with a unique RF station association: $ C = \{ (s_i, e_i, k_i) \} $ where $ s_1 < e_1 < s_2 < \dots < e_n $. (We note that multiple stations can fail simultaneously, for example after a fault in the sub-station. The uniqueness requirement is only to simplify the analysis metric and can be discarded during deployment.)

Generating the set of candidates for each station requires defining a function that takes the diagnostic data $ \{h_{k,t}\} $ and produces a set of candidates. We require this function to be computationally simple enough to run in real-time and produce no overlapping candidates. 
For the AMM status bit (0 - healthy, 1 - unhealthy), the function returns a candidate for every time range where the bit is continuously $ 1 $. (See section~\ref{sec:klystron_faults} for details.) Creating anomaly candidates from the real-valued AMPL signal is slightly more involved. LCLS has several automatic and operator-driven controls that constantly but slowly adjust the RF amplitudes. Therefore, we subtract the 3.5 minute rolling median value from the current AMPL value and threshold the absolute difference to create a new AMPL-based status bit. Using this computed status bit, we can recover anomaly candidates as with the AMM bit. The 3.5 minute rolling window duration was chosen to be substantially longer than short-duration dips, to prevent cases where the machine recovers on its own from strongly disturbing the median, but shorter than the time scale of machine drift. The threshold ($0.5\%$) was chosen to be deliberately tight (i.e., it will be prone to false positives) to flag subtle anomaly candidates, improving the sensitivity over the AMM-derived candidates. Additional details on how these diagnostic signals are converted into candidates can be found in Appendix~\ref{app:cand_gen}.

\subsection{\label{sec:fault_confirm}Anomaly Confirmation}
In the second stage, we use the BPM data to corroborate or reject anomaly candidates $ C_i = (s_i, e_i, k_i) $. We require this step to be:
\begin{itemize}
    \item \textit{Unsupervised}. There is a high volume of data, and manually labeling anomaly candidates is very time-intensive.
    \item \textit{Drift tolerant}. The BPM data distribution is constantly changing due to operator input and the automated feedback mechanisms.
    \item \textit{Jump sensitive}. Sensitive to sudden (and possibly short-lived) changes in BPM readings.
    \item \textit{Interpretable}. The anomaly confirmation should be easily interpreted by an operator and closely match their intuition.
    \item \textit{Real-time capable}. In order to deploy this in the operator control room, anomalies should be confirmed in `human-time,' (i.e., on the order of several seconds).
\end{itemize}

We use a simple strategy based on univariate median absolute deviation (MAD)~\cite{Hampel1974} and a geometric mean aggregation rule. This strategy requires no training, uses only two hyperparameters, and depends on only the last $\sim10$ seconds of data (making it drift tolerant, easily computed, and available in near real-time). Specifically, given a multivariate time series $ x_{s,t} $ for signal $ s $ at time $ t $, we define the anomaly score as follows:
\begin{enumerate}
    \item Run MAD on each BPM signal individually to get per-signal anomaly scores $ a_{s,t} $, using a 5 second rolling window. We use MAD as a robust z-score, where the mean and standard deviation estimates are replaced by robust median-based counterparts. The standard z-score, which measures the number of standard deviations above or below the mean, is quite sensitive to outlier behavior (e.g., the standard deviation will increase quickly and aggressively due to outlier data, lowering the z-scores of subsequent outlier data and thereby under-weighting the outlier severity).
    \item Aggregate across all signals to suppress beam noise and boost detection of beam energy changes. We use a geometric mean to more heavily penalize disagreement as compared to arithmetic mean. This yields a single anomaly time-series $ a_t $.
    \item Aggregate across several consecutive pulses. This boosts the score for sustained beam anomalies. Again we use a geometric mean.
    \item Threshold the maximum score in the anomaly window to determine if an anomaly has occurred.
\end{enumerate}
The full definition is given in Appendix~\ref{app:eneg_anom}. A key assumption here is that the beam is in a stable, non-anomalous condition prior to the RF station diagnostics identifying a potential problem. Therefore, the rolling windows used to define the anomaly score should capture the expected beam position and provide a meaningful baseline.

\section{\label{sec:results}Results}
We assess the performance of our method relative to the existing AMM-based detection system for the period stretching from November 2, 2020 to February 10, 2022. We only analyze periods of operation where the beam is being delivered to experiments with certain operational conditions to simplify data processing. We also remove time periods in which the BSA data-recording service itself had anomalous performance, which we identify using the BOCPD (Bayesian online changepoint detection)~\cite{Adams2007bayesian} algorithm. See Appendix~\ref{app:operational_beam} for full details on the data filtering.

As mentioned in Section~\ref{sec:fault_confirm}, labeling anomaly candidates is time-intensive. However, for the purposes of evaluating our approach, we hand-labeled the anomaly candidates generated during the start of our study window, specifically from November 2, 2020 to December 10, 2020. Labels follow the operators' process for identifying anomalous RF station behavior, selecting as anomalous any RF station diagnostic warnings that are nearly simultaneous with unusual changes in beam energy as measured by the BPMs.

\subsection{\label{sec:fault_ident_result}Improvement over Existing System}
During the labeled period, the AMM status bit identifies 1,237 anomaly candidates. However, we found that only 385 (31\%) coincided with visible drops in accelerator performance, meaning the vast majority of predicted anomalies are false positives. Additionally we will see later that the AMM bit misses a significant number of true anomalies as well. In sum, the AMM bit alone is an unreliable anomaly detector. 

The premise of this paper is that corroborating the AMM status bit (or other RF station diagnostics) with the BPM data using the procedure in Section~\ref{sec:method} improves the precision of the anomaly detection.  With corroboration, we reduce the number of false positives to $37$ ($3\%$), a factor of $20$ lower than using the AMM bit alone ($69\%$). At the same time, we still correctly identify 368 of the 385 anomalies present in the candidate set, meaning the corroboration only reduces the number of anomalies found (i.e., recall) by 4\%. The full confusion matrix is given in Table~\ref{tab:amm_cm}.

\begin{table}[htbp]
\caption{\label{tab:amm_cm}%
Confusion matrix for BPM corroboration of AMM-based candidates. 368 of the 385 anomalies are correctly confirmed; only 37 non-anomalies are mistakenly confirmed. Uses BPM threshold that maximizes the $F_1$ score. Note that the 17 false negatives consider only candidates selected by the AMM bit; true anomalies missed by the AMM bit itself are not considered here. 
}
\begin{tabular}{llrr}
\toprule
                                 &          & \multicolumn{2}{c}{True Class} \\
                                 &          & Anomaly      & No Anomaly        \\ \midrule
\multirow{2}{*}{Predicted Class} & Anomaly    & 368          & 37              \\
                                 & No Anomaly & 17            & 815            
\\\bottomrule                                 
\end{tabular}
\end{table}

\begin{figure}[htbp]
\centering
\includegraphics[width=\linewidth]{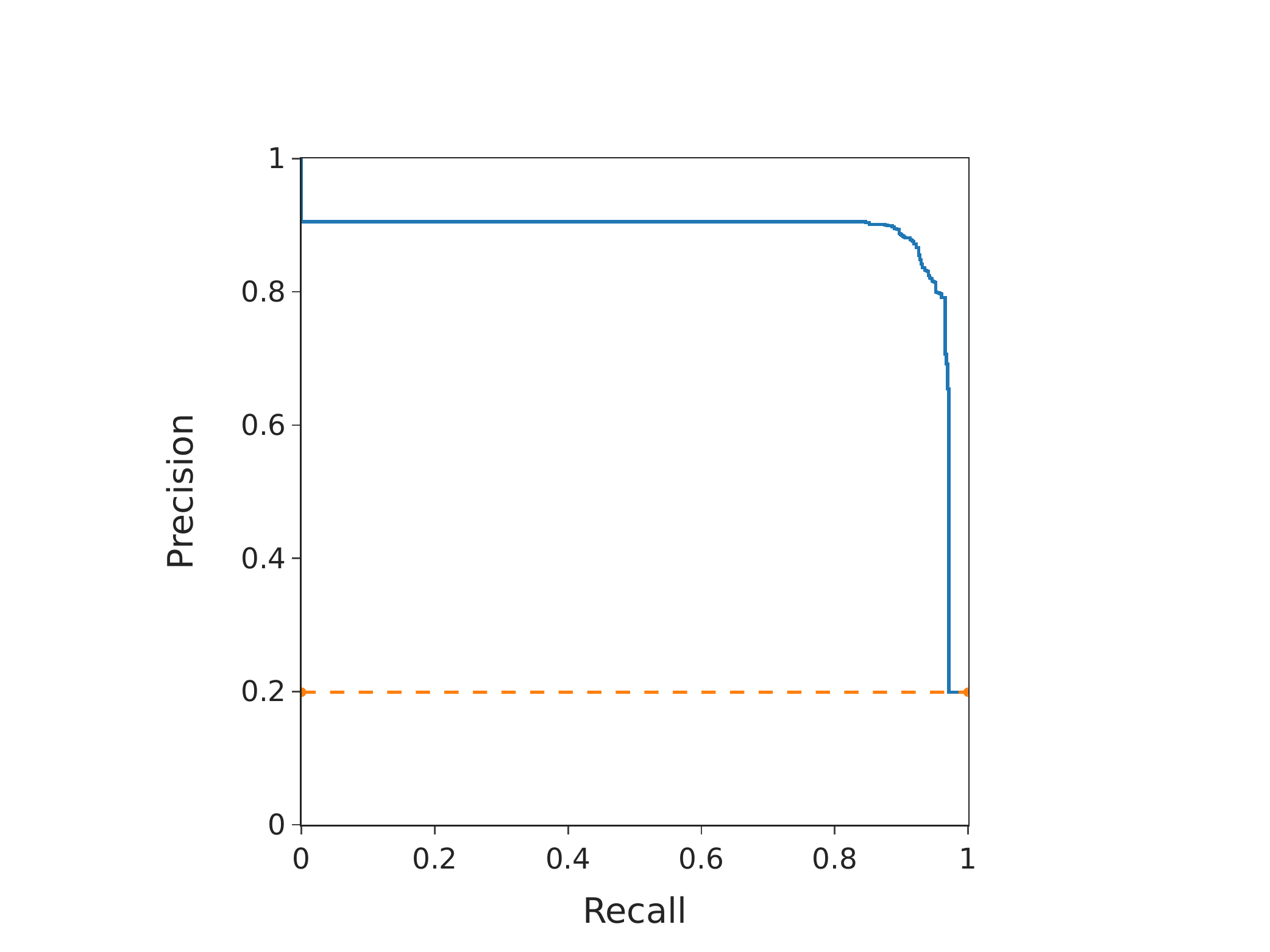}
\caption{\label{fig:ampl_pr} Interpolated precision-recall curve for the BPM threshold on AMPL-based candidates. The best $ F_1 $ score is $ 0.897 $, at a BPM anomaly score threshold of $ 2.848 $, corresponding to an accuracy of ~$96$\%. The dashed line is the precision-recall curve for a random classifier. Note that recall here does not include anomalies missed by the AMPL threshold, and is only an upper limit on the recall for the full algorithm.}
\end{figure}

Since we suspected that the AMM bit was missing anomalies, we also investigated using the underlying AMPL signal to improve the recall of the RF station anomaly detection. With a permissive threshold of just $0.5\%$, the AMPL signal identifies $2,845$ anomaly candidates during the labeled period, more than twice as many identified by the AMM bit. Of these we find $553$ ($19.4\%$) coincide with drops in accelerator performance according to the BPM data. Using the underlying AMPL signal therefore identifies at least 168 anomalies that are missed by AMM, increasing recall by $44\%$. 
Note that both AMM and AMPL likely fail to identify true anomalies as candidates due to the bandwidth and noise issues mentioned in Section~\ref{sec:klystron_faults}, leading to additional false negatives. Future work will aim to predict anomalies solely from full beam-rate BPM data to capture additional anomalies (see Section~\ref{sec:dataset}). The full confusion matrix and interpolated precision-recall (PR) curve for the AMPL signal are shown in Table~\ref{tab:ampl_cm} and Figure~\ref{fig:ampl_pr} respectively. To obtain the PR curve and confusion matrix, we vary only the threshold on the BPM-based anomaly score from Section~\ref{sec:fault_confirm}, keeping the $0.5\%$ permissive AMPL deviation threshold fixed.

The precision and the number of anomalies found with all three methods are given in Table \ref{tab:pr}. (Note that raw number of anomalies are given rather than recall, because there are likely additional anomalies that are undiscovered by any of the existing methods.) The addition of the BPM corroboration to the AMM bit results in only a small drop in the number of anomalies found, while greatly improving precision. Switching the trigger to the AMPL signal significantly improves the number of anomalies found at the expense of a small drop in precision. 
\begin{table}[htbp]
\caption{\label{tab:ampl_cm}%
Confusion matrix for BPM corroboration of AMPL-based candidates. 504 of the 553 anomalies are correctly confirmed; only 68 non-anomalies are mistakenly confirmed. Uses BPM threshold that maximizes the $F_1$ score. Note that the 49 false negatives consider only candidates selected by the AMPL signal; true anomalies missed by the AMPL signal itself are not considered here. 
}
\begin{tabular}{llrr}
\toprule
                                 &          & \multicolumn{2}{c}{True Class} \\
                                 &          & Anomaly      & No Anomaly        \\ \midrule
\multirow{2}{*}{Predicted Class} & Anomaly    & 504          & 68              \\
                                 & No Anomaly & 49           & 2224
\\\bottomrule
\end{tabular}
\end{table}

\begin{table}[htbp]
\caption{\label{tab:pr}%
Precision and number of anomalies found using the AMM bit alone, AMM bit corroborated by BPMs, and AMPL corroborated by BPMs. The corroborated data uses the BPM thresholds that maximize the $F_1$ score. 
}
\begin{tabular}{lcc}
\toprule
    & Precision & Anomalies found  \\\midrule
AMM-only    & 0.31      &       385      \\
AMM+BPM     & 0.91      &       368      \\
AMPL+BPM    & 0.88      &       504      \\\bottomrule
\end{tabular}
\end{table}

Both the AMM and AMPL signals are evidently noisy and inaccurate to differing degrees. By combining these signals with the anomaly detection algorithm on the full-rate BPM data, we are able to significantly improve detection accuracy by rejecting numerous false positives. Furthermore, using the AMPL signal in addition to the AMM status bit reduces the number of missed anomalies (false negatives). In particular, of the $ 385 $ true anomalies detected by the AMM signal, the AMPL signal flagged $ 305 $ of them while only missing $ 13 $ anomalies. The remaining $ 67 $ anomalies were given a different or multiple attribution according to the AMPL signal. Of the $ 553 $ true anomalies detected by the AMPL signal, the AMM signal flagged $ 307 $ of them but missed $ 245 $; there was $ 1 $ anomaly that AMPL attributed to KLYS:LI29:11 but AMM attributed it to KLYS:LI27:41.

\begin{figure*}[htbp]
    \centering
    \includegraphics[width=\linewidth]{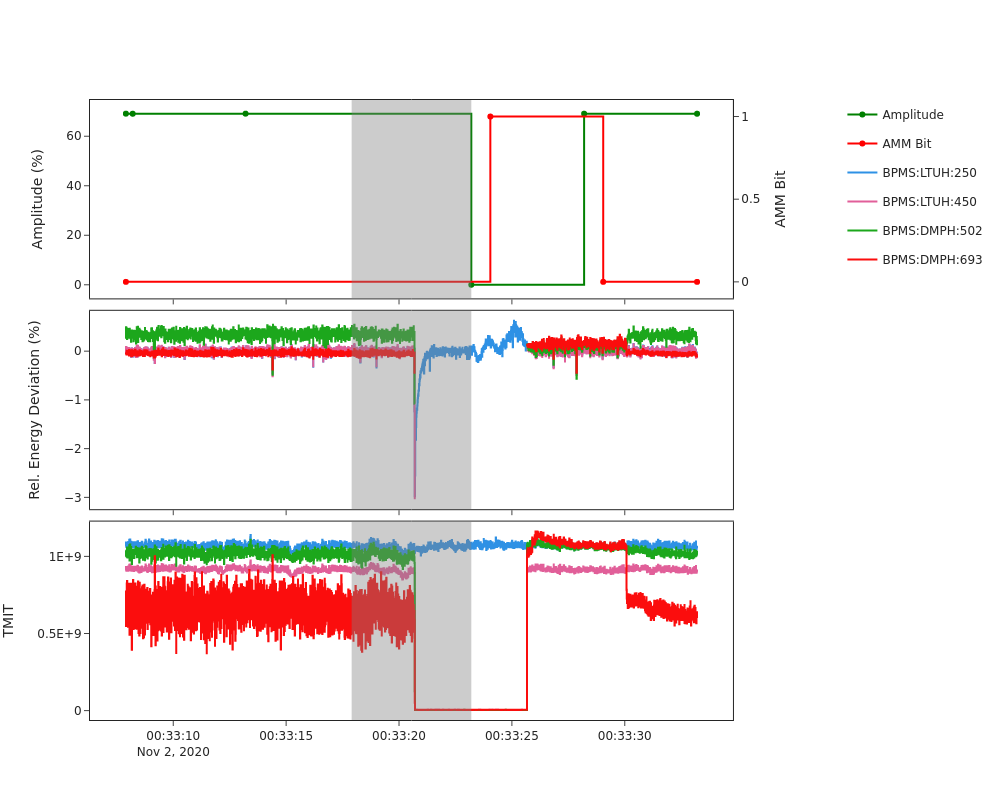}
    \caption{Example of an RF station fault at KLYS:LI29:11. The top plot shows the klystron diagnostic data (both the AMM status bit and the AMPL signal). The middle plot shows the relative energy deviation (calculated from the beam position in the dispersive plane). The bottom plot shows the TMIT readings (which is the approximate number of electrons, subject to some calibration). The grey region defines the implied anomaly window, due to the delayed reporting of the klystron health information. This extreme example shows the AMPL drop nearly to 0, which causes the beam to be lost. The missing pulses in the energy deviation plot are the result of there being no beam to measure.}
    \label{fig:catastrophic_fault} 
\end{figure*}

\begin{figure*}[htbp]
    \centering
    \includegraphics[width=\linewidth]{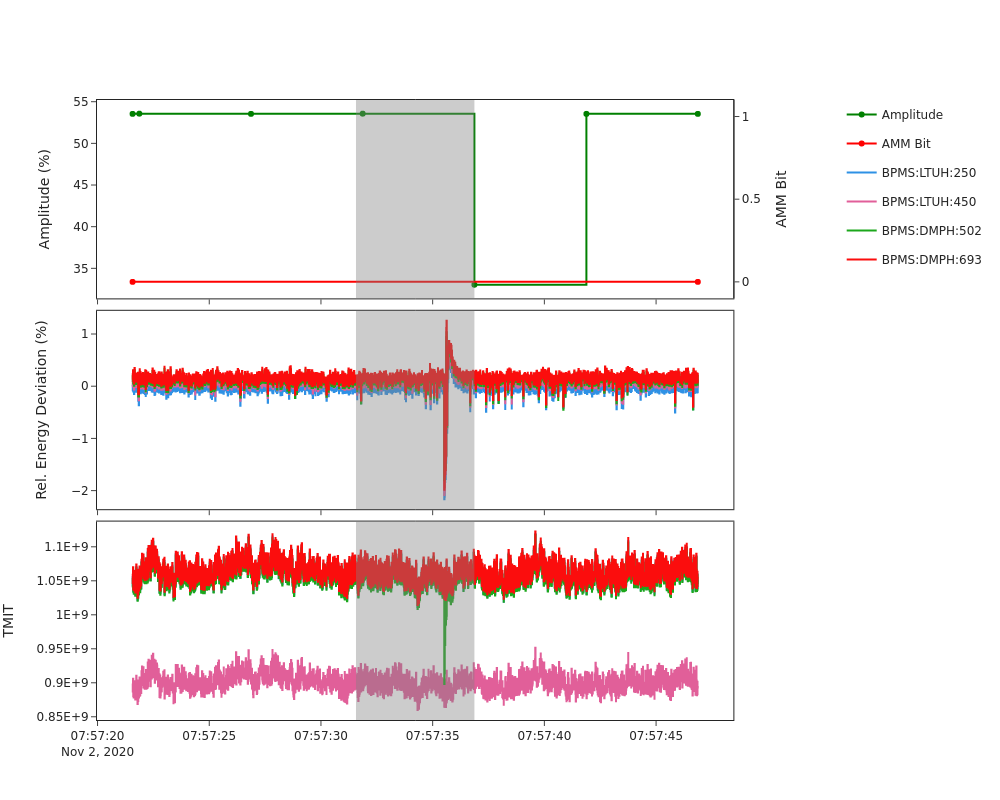}
    \caption{Example of a sustained anomaly at KLYS:LI28:11. The top plot shows the klystron diagnostic data (both the AMM status bit and the AMPL signal). The middle plot shows the relative energy deviation (calculated from the beam position in the dispersive plane), with a brief but large deviation followed by a recovery period. The bottom plot shows the TMIT readings (which is the approximate number of electrons, subject to some calibration). The grey region defines the implied anomaly window, due to the delayed reporting of the klystron health information. Note this anomaly is missed by AMM.}
    \label{fig:sustained_fault}
\end{figure*}

As a byproduct of our analysis, during hand labeling of the anomaly candidates, we discovered a natural grouping of the candidates into two types:
\begin{enumerate}
    \item \emph{Faults}: beam is lost (TMIT goes to 0). Figure~\ref{fig:catastrophic_fault} shows an example.
    \item \emph{Sustained anomalies}: initial large deviation followed by a recovery period lasting several hundred pulses. Figure~\ref{fig:sustained_fault} shows an example, in which AMM misses the anomaly.
\end{enumerate}

Of the $553$ anomalies detected by combining the AMPL signal and BPM data, $ 217 $ were \textit{Faults} and $ 336 $ were \textit{Sustained anomalies}. The majority of the anomalies, including both examples (Figures~\ref{fig:catastrophic_fault},\ref{fig:sustained_fault}), are short lived, lasting less than $\sim10$ seconds. Human operators are likely to miss these anomalies, as by the time they notice a degradation in beam performance the problem is likely resolved and the klystron health signal reverted to normal. Particularly in the case of \textit{Sustained anomalies}, human operators might not even notice the moderate performance degradation, despite the impact on FEL energy and therefore user experiments.

\subsection{\label{sec:full_anom_result}Anomalies at LCLS}

\begin{figure*}[t]
\centering
\includegraphics[width=\linewidth]{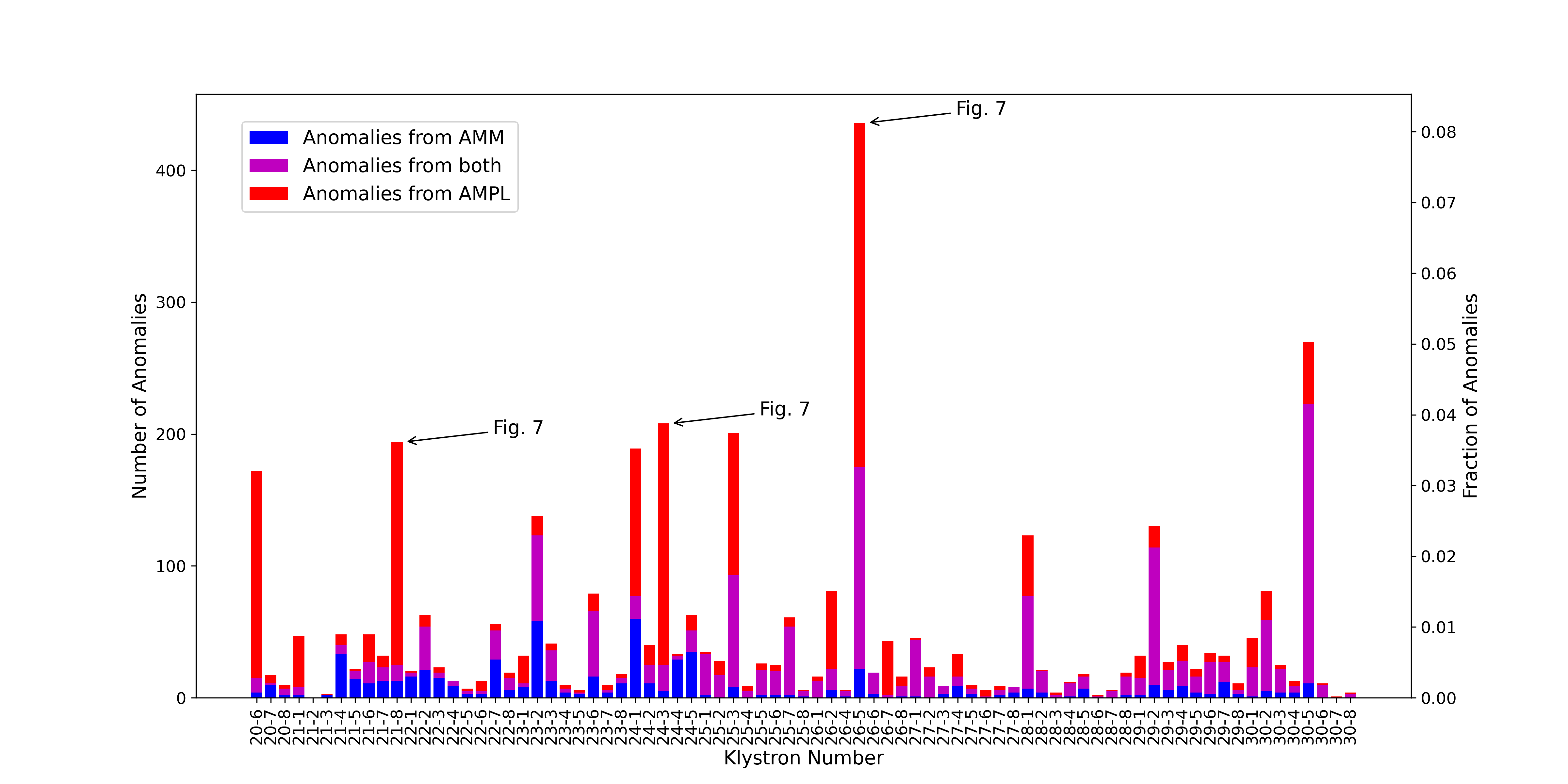}
\caption{\label{fig:lcls_anomalies} Count of anomaly candidates identified by either the AMM bit, the AMPL signal, or both that were labeled anomalies using our anomaly detection routine on the BPM data.  The red part of the bar is anomaly candidates detected only by the AMPL signal, the blue part of the bar is anomaly candidates detected only by the AMM bit, and the magenta part of the bar is anomaly candidates that were detected by both.   The klystrons are enumerated along the LCLS accelerator from low energy (left) to high energy (right).}
\end{figure*}


We now apply the same algorithm in an unsupervised manner to the entire dataset to search for anomalies.  After applying the criteria for data selection at the beginning of Sec. \ref{sec:results}, we have approximately 80 days worth of data to survey.  Over this time period, we find 29,739 anomaly candidates in the RF station diagnostic data: 3,615 from the AMM bit alone, 22,570 from the AMPL signal alone, and 1,777 found by both.  Of these 6,529 (21.9\%) exceed the BPM anomaly threshold found in the previous section and are labeled as anomalous. Of the more restrictive AMM candidates, 47.5\% are labeled as anomalous, while only 16.3\% of the more permissive AMPL signal candidates are labeled as anomalous. Fig \ref{fig:lcls_anomalies} shows that anomalies are not distributed equally among the the RF stations; in particular, the AMPL signal usually identifies many more anomalies in the RF stations that experienced the most anomalies. The top 5 RF stations (of 82) account for \(\sim36\)\% of the identified anomalies. 

It is possible that the BPM data contains anomaly-like behavior even in the absence of an RF station anomaly, which would lead to corroboration of false positives. To investigate the false positive rate, we take 6,453 random samples (approximately half of a day) of the BPM data independent of the RF station diagnostics.  All samples of the BPM data are of the same length as the candidate samples.  We find 176$\pm 13$ (2.7$\pm0.2$\%) samples exceed the anomaly threshold and are labeled anomalies despite no apparent anomaly in the RF station data (the 1-sigma confidence interval assumes the random samples are i.i.d.\ Bernoulli random variables). Taking the conservative assumption that all events found in the random stream are false positives sets an upper limit on the false-positive rate.  

A secondary benefit of an automated detection scheme is the ability to log large numbers of events to track performance, predict future failures, or enable later analysis. At present, RF station anomalies are logged through the manual CATER system \cite{slac_cater}. The CATER reports are limited by the need for operator attention; operators may not notice a transient anomaly, or may not have the time to manually log an event, and there is no way for one or two operators to watch thousands of potential signals. Moreover, the CATER system does not try to quantify the anomaly rate for any particular subsystem.  

\begin{figure}[htbp]
\centering
\includegraphics[width=\linewidth]{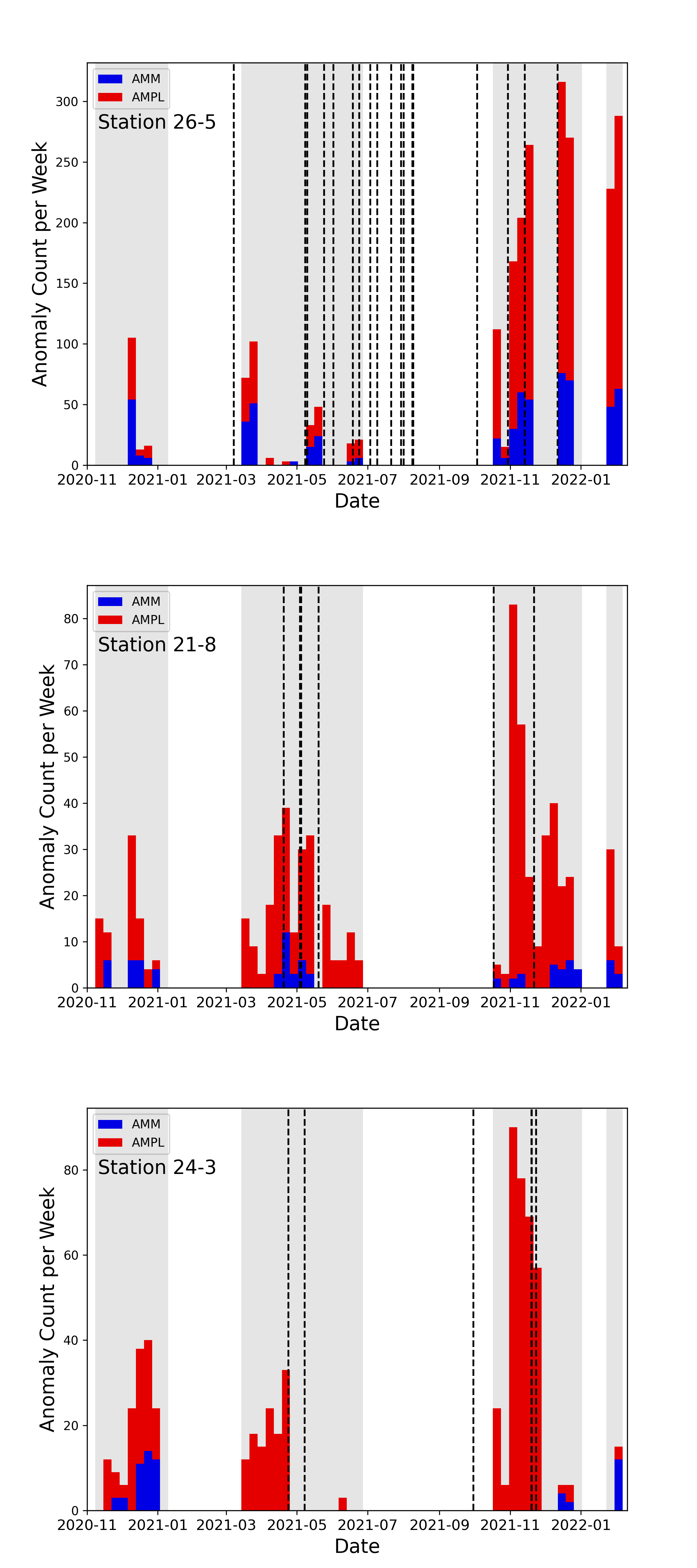}
\caption{\label{fig:klys_vert} Count of the number of anomalies detected by AMPL (red bars), the AMM bit (blue bars) and problems reported in CATER (black dashed lines) for various stations. The anomaly counts have been binned by week to produce the histograms.  The shaded regions show weeks where the BPM data was available through the BSA service.
(top) Station 26-5.  For this station both the AMPL and AMM bit signals detect anomalies at approximately the same time.
(middle) Station 21-8. AMPL detects anomalous activity more often than the AMM bit does.
(bottom) Station 24-3. AMPL detects anomalous activity in April and November 2021 that the AMM bit misses entirely.
}
\end{figure}

To assess our automated approach, we compare the automated logging to CATER events during the studied time period. In total there were 597 requests for work made to the groups charged with maintaining the RF stations.  444 of these reference a specific RF station, compared to 6,529 for the automated approach despite the BSA service being active for only a subset of the time period. All 82 stations studied in this work are referenced at least once.  Figure \ref{fig:klys_vert} show predictions for three of the stations showing the most anomalies: 26-5, 24-3, 21-8.

Station 26-5 had the most anomalies of any station.  Figure \ref{fig:klys_vert} (top) shows a histogram of the anomalies (binned by week) and the times at which problems with the RF station were reported through CATER.  Both the AMM bit and the AMPL signal capture events. The performance on the data from stations 25-3 and 30-5 (not shown) was similar. For stations 21-8 and 24-3, the importance of the AMPL signal specifically is more apparent (see Fig. \ref{fig:klys_vert} middle and bottom). For 24-3, there are two periods, in April and November 2021, where our algorithm detects anomalies that were totally missed by the AMM bit. These periods of high anomaly counts were followed by several CATER events. Station 21-8 shows more overlap (in time) of the anomalies identified using the AMM bit and the AMPL signal, but there are still periods of anomalies missed entirely by the AMM bit. 

\subsection{\label{sec:dataset}RF Station Anomaly Dataset and Classifier}
For the labeled study period, we have compiled a dataset of the RF station diagnostic data and the BPM signals, alongside the hand labels, as a public anomaly detection dataset. The dataset consists of two HDF5 files (one for AMM and one for AMPL) containing the raw data, two CSV files containing information about the candidates, and two CSV files containing the hand labels. We also include in the HDF5 files random samples of beam operation to serve as negative or background examples. A full description of the files can be found in Appendix~\ref{app:dataset}. The files can be accessed at \url{https://www.slac.stanford.edu/grp/ad/ard/rfanom/rfanom.html}.

While our study focused on unsupervised anomaly detection, the existence of a labeled dataset opens the possibility to train more powerful supervised models, including deep neural networks (DNNs). For example, we have only looked at the most direct RF station and BPM signals in the amplitude and beam energy respectively. Future work could extend to more subtle diagnostics, such as phase information for the RF station or non-dispersive BPMs. RF phase information is expected to reveal additional anomalies not observed solely through amplitude variation, but the signal was too complex to be analyzed with an unsupervised method. A DNN may be more successful. Similarly, non-dispersive BPMs located throughout the accelerator are both full beam-rate and have RF station specificity, and may be able to sidestep the bandwidth limitations of the RF station diagnostics entirely. While the non-dispersive BPMs are sensitive to many other types of anomalies that would confuse our unsupervised approach, a DNN trained on a labeled dataset may be capable of distinguishing anomaly categories and identifying the faulty RF station. As a brief demonstration, we trained a deep binary classifier using the raw BPM signals as input. The experimental setup is described in Appendix~\ref{app:bin_classifier}. On the $ 20\% $ held-out test set, which contains $ 569 $ candidates, our trained model achieved an F$1$ score of $ 0.88 $, correctly identifying $95\%$, showing the DNN is at least as accurate as the algorithm presented in Section~\ref{sec:fault_confirm}. We present the full precision-recall curve in Appendix~\ref{app:bin_classifier}. Further improvement will require extension of the dataset to include RF phase diagnostics or non-dispersive BPMs. We leave the problem of identifying RF station performance changes solely from BPM data to future work. We emphasize that this proof-of-concept DNN is not targeted towards deployment in the operation room but is mostly a suggestion of the potential future work.

\section{\label{sec:conclusion}Conclusion and Discussion}
This work describes a beam-based approach to automating the detection of anomalous behavior at LCLS. We focus on a particularly common failure mode, RF station anomalies, and improve upon the existing diagnostic system. We have developed a fully automated system that combines slow-updating, inaccurate RF station diagnostic data with the full-rate BPM data. By corroborating the diagnostic data with measurements of the beam, we are able to filter out more than $95\%$ of the false positives signaled by the RF station diagnostic data. Moreover, by using the underlying AMPL signal, in addition to the current AMM status bit, we detect significantly more RF station drops and faults, reducing the number of anomalies missed (false negatives). On a hand-labeled dataset, our approach improved precision and number of events found from $0.31$ and $385$ for the raw AMM bit to $0.88$ and $504$ for AMPL corroborated by the beam. On an extended unlabeled dataset, our algorithm found a total of $ 6529 $ events, compared to $ 597 $ CATER events logged during the same time-period. The proposed algorithm will give operators a higher degree of confidence that RF station faults are being detected and the responsible station identified, and provides RF technicians with a detailed history of faults to guide repairs.


The need to combine expert knowledge with anomaly detection systems will only grow as accelerator systems become more complex. Moreover, as the data volume grows, data collection cannot always be at the full system rate, forcing some signals to be recorded at slower rates. The beam-based method presented, leveraging high-fidelity diagnostics on the beam to corroborate low-fidelity sub-system diagnostics, may find common use in other accelerator anomaly prediction tasks. Future work will expand the existing dataset and can serve as a test-bed for development of new beam-based anomaly detection algorithms for accelerators. We also plan to detect and classify RF station anomalies using data from both dispersive and non-dispersive BPMs, as different RF stations will affect the beam to different degrees and in different manners.

\begin{acknowledgments}
Use of the LINAC Coherent Light Source (LCLS), SLAC National Accelerator Laboratory, is supported by the U.S. Department of Energy, Office of Science, Office of Basic Energy Sciences under Contract No.\ DE-AC02-76SF00515.
\end{acknowledgments}

\appendix

\section{\label{app:cand_gen}Creating Candidates}
In practice, the AMM status bit is a noisy and unwieldy signal, so creating candidates from it is not as trivial as taking consecutive regions where the status bit is $ 1 $. The AMM status bit is itself the result of applying a operator-configurable threshold to internal klystron diagnostic information. Unfortunately, this threshold can often be misconfigured, leading to $ h_{k, AMM, t} = 1 $ for implausibly long periods of time. This affects the candidate identification step by technically overlapping with many other possible candidates, which for the sake of interpretability, we would then drop. To counteract this, and attempt at a fairer comparison with AMPL, we will ignore $ h_{k, AMM, t} $ if it exceeds 10\% unhealthy over a period of otherwise apparently normal beam operation. This trouble in generating anomaly candidates from AMM further supports the conclusion that AMPL is a better diagnostic for RF station health.

Also, each klystron is not constantly in use, with at least a couple held in reserve for operators to swap out malfunctioning klystrons for healthy ones. Therefore, before assessing the diagnostic signal (AMM or AMPL), we need to first check if the klystron is in use. This information is stored in a different PV in the EPICS Archiver, which we fetch alongside the diagnostic signal itself.

Lastly, the klystron signals are only recorded in the EPICS Archiver infrequently ($\sim 0.2$Hz for AMM and AMPL) and only when the value changes. This has the effect of timestamping the AMM and AMPL data in the Archiver with a delay possibly as large as the inverse of the logging rate. Therefore, we say AMM or AMPL imply a candidate window in the $\sim 5$ seconds before the timestamp actually stored in the Archiver. Also, when we take rolling windows in creating the AMPL-based candidates, we need to carefully weight the data due to the non-uniform sampling in time.

\section{\label{app:eneg_anom}Anomaly Confirmation Algorithm}
Given the multivariate time series BPM data $ x_{s,t} $ for signal $ s $ at time $ t $, we define the per-signal anomaly score as $ a_{s,t} = \text{MAD}(x_{s, t-2l}, \dots, x_{s,t}) $ where $ l $ is the lagging rolling window used to compute the median and median-absolute deviation of the signal $ x_{s, t} $. Specifically, we estimate $ \hat{\mu}_{s, t} = \text{median}( x_{s, t-l}, \dots, x_{s, t-1} ) $ and $ \hat{\sigma}_{s, t} = k \, \text{median}( | x_{s, t-l} - \hat{\mu}_{s, t-l} |, \dots, | x_{s, t-1} - \hat{\mu}_{s, t-1} | ) $. The extra factor $ k $ ensures that $ \hat{\sigma}_{s, t} $ is a consistent estimator of the true standard deviation, where the underlying data is normally distributed. Rigorously, $ k = 1/\Phi^{-1}(3/4) \approx 1.4286 $ (where $ \Phi(x) $ is the CDF of a normal distribution)~\cite{Hampel1974}. Thus, $ \text{MAD}(x_{s, t-l}, \dots, x_{s,t}) = \hat{\sigma}_{s,t}^{-1} \left( x_{s,t} - \hat{\mu}_{s,t} \right) $.

Now, we want to aggregate into a single anomaly score $ a_t $. Recall that we are using both the dispersive X/Y signals and the TMIT signals from the 4 dispersive BPMs at the end of the beamline. Unfortunately, if the anomaly is severe enough to cause loss of beam at a BPM (TMIT lower than $1e8$), the X/Y reading for that pulse is simply repeated from the last known value. This can make the dispersive reading appear perfectly normal. Therefore, we define the anomaly score for each BPM as the anomaly score associated with the TMIT signal if the beam is lost (as the TMIT anomaly score will be large); otherwise, we define it to be the anomaly score associated with the dispersive signal.

Given an anomaly score for each BPM then, we finally create a single anomaly score using a geometric mean rule: $ a_t = \left( \prod_s a_{s,t} \right)^{1/n} $. We further aggregate across $ 10 $ consecutive pulses to get the final score $$ a_{AGG,t} = \left( \prod_{i=0}^{9} a_{s,t-i} \right)^{1/10} $$ to boost the score for sustained beam anomalies. Lastly, each candidate (as identified in~\ref{sec:cand_gen}) suggests an anomaly might have occurred during some time period from start time $ s $ to end time $ e $. We say the BPM data confirms the anomaly if $$\max_{s \leq t \leq e} a_{AGG,t} \geq \tau $$ for a configurable threshold $ \tau $.

\section{\label{app:operational_beam}Beam and BSA status filters}
During our study period from November 2, 2020 to February 10, 2022 ($457$ days), we applied a series of data filters, ultimately yielding $\sim80$ days of data. The aim of these filters is to ensure the beam is in a steady operation mode within normal operational bounds and that the recorded BPM data is quality. We also restrict this analysis to hard X-ray operation, because the hard and soft lines use different BPMs; this restriction can be lifted in the future with additional data processing work.

We start by imposing a number of checks on the beam state:
\begin{enumerate}
    \item Does the beam have standard charge at the injector? We require $ \text{BPMS:IN20:221:TMITCUH} > 0.5e9$ and be logged every second in the EPICS Archiver. Lower charge or the charge not being logged can indicate the beam is not being operated in a standard operational mode.
    \item Is the beam stopper being used? We require $ \text{STPR:BSYH:2:STD2\_IN\_A} = 0 $, indicating the beam stopper is out.
    \item Is the beam rate 120Hz? We require $ \text{IOC:BSY0:MP01:PC\_RATE} = 8$ .
    \item Is the entire beam being delivered to the hard X-ray line? We require $ \text{IOC:IN20:EV01:RG02\_ACTRATE} = 10 $.
\end{enumerate}
We do allow temporary ($<90$ second) violations of this conditions to not disallow short periods of ``non-standard'' beam operation caused by automatic feedback or protection-based control mechanisms. This reduces the $457$ day study period to $\sim153$ days of beam operation ($\sim103$ days are removed due to charge violations, $\sim76$ days are removed due to beam delivery violations, and $\sim125$ days are removed due to both).

Separately, we check if the data from the BSA service~\cite{kim_bsa} exists during these periods of beam operation and was recorded correctly. Due to a number of engineering issues, significant portions of our study window do not have good quality data. The BSA service records over $1000$ PVs synchronously at the beam operational rate into hour-long HDF5 data files. To verify the quality of the data, we check the following:
\begin{enumerate}
    \item Do the data files exist? Many data files were simply never written due to issues with the BSA service.
    \item Can we read the data from the file? Some data files are corrupted and cannot be read.
    \item Was real data recorded? We drop any data where NaNs are recorded.
    \item Was the data recorded at the correct rate? We expect data to be recorded at $120$ Hz. This check was somewhat challenging, so we discuss it at more length below.
\end{enumerate}
Of the $457$ day study period, $231$ days (just over $50\%$) must be dropped due to a violation of one of these conditions. The primary reason was missing data files: roughly $ 154 $ days of data were not recorded. The majority of the remaining $72$ invalidated days were due to an invalid record rate.

Unfortunately, the rate recording issues varied widely and required additional work to discover. Let $ t_i $ be the timestamp for the $i$th row of data. Since the standard operational rate of LCLS is $120$Hz, we expect $ \Delta_i = t_i - t_{i-1} \approx 1/120 $ s. We first define a timestamp mask $ m_i = \mathbf{1}\left( \left| \Delta_k - 1/120 \right| \leq 2.5e-4 \right) $; this defines whether the current row is recorded at the correct rate with respect to the previous row, up to a tolerance. To identify time periods with bad data recording, we employed the BOCPD (Bayesian online changepoint detection) ~\cite{Adams2007bayesian} algorithm. Using a constant hazard $ \lambda = 1e5 $ and a binomial likelihood with prior $ p = 0.9 $ with strength (pseudo-observations) $ 2 $, we ran the BOCPD algorithm on the mask $ m $ signal. We define a changepoint at time $ i $ if the probability of having a run of length $ 100 $ at time $ i + 100 $ exceeds $ 10\% $. This yields a list of time periods whose mask behavior is statistically distinct from their neighbors. Finally then, we drop time periods where the average mask value was below $ 85\%$. All of these hyperparameters were chosen on a best effort/empirical basis to remove known periods of invalid rate recording. We set the seemingly permissive $ 85\% $ threshold because the rate recording issue can in fact often be repaired. The most common observed rate failure mode was a double logging issue, where $ \Delta_i \approx 1/120 $, $ \Delta_{i+1} \approx 0 $, and the $i+1$ data row was an exact duplicate of the previous row $ i $. Therefore, we simply repaired these periods by dropping the offending duplicate rows.

To further refine our search period to periods of constant beam delivery, we only analyze periods of operation where all of the above are continuously satisfied for at least $ 5 $ minutes.  Combined together, these conditions reduced our study period from $457$ days to only $\sim80$ days of operation that could be analyzed.

Lastly, as an intricacy of LCLS, there are certain operation modes that forcibly kick pulses at a configurable cadence out of the beamline. This has the effect of making the TMIT reading at our dispersive BPMs crash towards $ 0 $. Without additional handling, these kicks appear to an observer as single pulse anomalies, despite this phenomenon being a deliberate, operator-controlled behavior. Therefore, we apply a pre-processing step to remove these kicked pulses from our data.

\section{\label{app:dataset}Dataset Details}
The RF station anomaly dataset is composed of several files. For both AMPL and AMM, there are three files: ``klys\_anom\_dset\_\{type\}.h5'', ``candidates\_\{type\}.csv'', and ``labels\_\{type\}.csv''.

The HDF5 file has two top level HDF5 groups, one for \textit{candidates} and one for \textit{samples}, corresponding to the anomaly candidates and random samples of the beam operation respectively. Each top-level group is composed of a number of subgroups corresponding to the individual examples. The subgroups are named according to the end time of the anomaly candidate (or randomly sampled window); the time is stored as nanoseconds since epoch. Each subgroup contains two datasets called \textit{health} and \textit{bpm}. Each dataset is written as a 2D array of data, where the rows correspond to individual measurements and the columns correspond to the signal name. The timestamps of the rows are stored as the attribute \textit{index} on the dataset; the values denote the number of nanoseconds past epoch. The column names are stored as the attribute \textit{columns} on the dataset. The subgroups within \textit{candidates} additionally have a group attribute called \textit{klys}, which lists the RF station associated with the candidate. 

The \textit{health} dataset stores the RF diagnostic data of the given type over a period of $ 3.5 $ minutes (which corresponds to the look-back duration of the algorithm defined in~\ref{sec:cand_gen} for AMPL). It will have shape $ (n, 82) $ where $ n $ is the number of measurements and $ 82 $ is the number of RF stations. Because the RF station diagnostic data is not beam synchronous, the reporting timestamps are not aligned with one another nor the beam, meaning $ n $ can vary from dataset to dataset. In particular, it can be assumed that the AMM and AMPL diagnostic data can be reported with a delay up to roughly $ 5 $ seconds. All of the NaN values correspond to no new data for that RF station at that timestamp, and the actual value can be assumed to be equal to the last known value. If there is no good last value, then the RF station was not in use during that period of time.

The \textit{bpm} dataset stores the BPM data over a period of roughly $ 17 $ seconds (which corresponds to the look-back duration of the algorithm defined in~\ref{sec:fault_confirm}). It will have shape $ (n, 8) $ where $ n $ is the number of pulses and $ 8 $ is the number of signals ($ 2 $ signals for each of the $ 4 $ dispersive BPMs of interest). The number of pulses $ n $ can vary due to the rate recording and pulse-kicking behaviors mentioned in~\ref{app:operational_beam}. This BPM data was additionally modified from the original raw data by several transforms. First, if the BPM's TMIT value was below $ 1e8 $, there was not enough beam to record a reliable value of the dispersion, so the raw data merely repeats the last known value. This often erroneously repeats a ``normal'' value, so we artificially set the dispersive reading to $ 100 $ (well outside its feasible range) when the minimum TMIT condition is violated. Second, we apply a dispersive scaling to each BPM to place it on the order of beam energy loss. The scalings are as follows: (BPMS:LTUH:250, 0.125), (BPMS:LTUH:450, -0.125), (BPMS:DMPH:502, 0.964), and (BPMS:DMPH:693, 0.469).

The candidates CSV file lists all of the candidates and has the following columns:
\begin{enumerate}
    \item start: Earliest time anomaly could have occurred
    \item end: Latest time anomaly could have occurred
    \item klys: Associated RF station
    \item source: Reporting diagnostic signal (AMM or AMPL)
    \item corroborate: Can take several different values, depending on the what the other diagnostic signal (AMM if considering AMPL candidates and vice versa) reports
        \begin{enumerate}
            \item Missing: other diagnostic signal did not flag this candidate
            \item Reinforced: other diagnostic signal flagged the candidate and supports the attribution
            \item Disagree: other diagnostic signal flagged the candidate but disagrees on the RF station attribution
        \end{enumerate}
    \item corr\_anomaly\_list: Lists the associated RF stations and the reporting diagnostic source
\end{enumerate}

The labels CSV file gives the hand-labels for the candidates. Like the candidates file, it has start and end columns. It has two additional columns: 
\begin{enumerate}
    \item is\_anom: Boolean indicating whether the candidate is an anomaly
    \item anom\_type: Either ``f'' for \textit{Fault}, ``s'' for \textit{Sustained anomaly}, or blank for no anomaly
\end{enumerate}

\section{\label{app:bin_classifier}Classifier Experimental Setup}

\begin{figure}[htbp]
    \centering
    \begin{tikzpicture}[node distance=1.1cm]
        \node (data_in) [] {Input};
        \node (conv1) [nn, below of=data_in] {Dilated 1D Convolution + BN + ReLU};
        \node (conv2) [nn, below of=conv1] {Dilated 1D Convolution + BN + ReLU};
        \node (conv3) [nn, below of=conv2] {Dilated 1D Convolution + BN + ReLU};
        \node (fc1) [nn, below of=conv3] {Fully-connected + BN + ReLU};
        \node (fc2) [nn, below of=fc1] {Fully-connected + Sigmoid};
        \node (pred) [below of=fc2] {Predicted Label};

        \draw [arrow] (data_in) -- (conv1) node [pos=0.5, right] {(8,720)};
        \draw [arrow] (conv1) -- (conv2) node [pos=0.5, right] {(32,176)};
        \draw [arrow] (conv2) -- (conv3) node [pos=0.5, right] {(64,40)};
        \draw [arrow] (conv3) -- (fc1) node [pos=0.5, right] {768};
        \draw [arrow] (fc1) -- (fc2) node [pos=0.5, right] {128};
        \draw [arrow] (fc2) -- (pred) node [pos=0.5, right] {1};
    \end{tikzpicture}
    \caption{\label{fig:dnn_arch} Model architecture of the demonstration DNN, showing the input size to each layer.}
\end{figure}
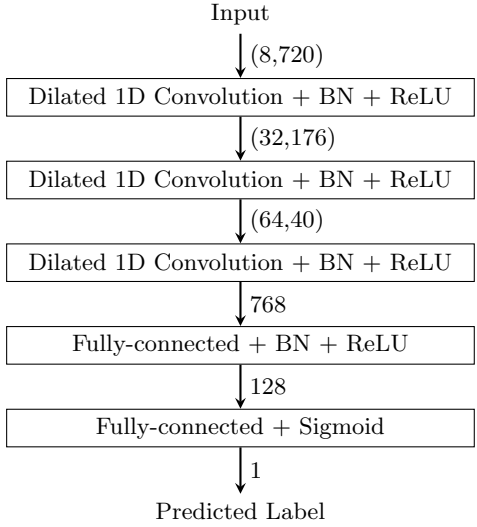

\begin{figure}[htbp]
    \centering
    \includegraphics[width=\linewidth]{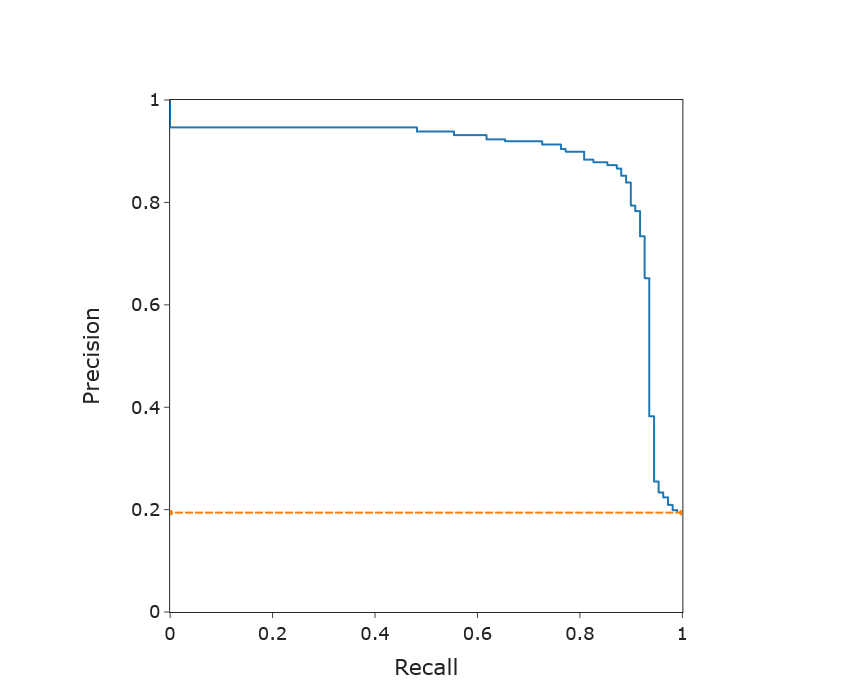}
    \caption{\label{fig:dnn_prc}Interpolated precision-recall curve for the DNN classifier on our new RF station anomaly dataset. The dashed line is the precision-recall curve for a random classifier.}
\end{figure}

We trained a deep neural network on the BPM data and labels described in Appendix~\ref{app:dataset}. We use the last $ 720 $ pulses before the candidate's end time to create a constant input size of $ (8, 720) $ to our network, which we defined using PyTorch~\cite{paszke_2019} and initialized using Kaiming's uniform distribution~\cite{HeZRS15}. As shown in Figure~\ref{fig:dnn_arch}, our network consists of $ 3 $ 1D convolutions with kernel size $ 8 $, padding $ 1 $, dilation $ 3 $, stride $ 4 $, and out channels of $ 32, 64, 128 $ respectively, followed by two FC layers of output size $ 128 $ and $ 1 $. All layers but the final layer are followed by a 1D batch normalization and ReLU activation.
We split the dataset into a train and test set with a $80\%/20\%$ split and normalize the data using the median and MAD of the train data. We use a binary cross entropy loss with positive weighting equal to $ 4.15 $, since the candidate set is only $ 19.4\% $ anomalous. We use the Adam optimizer~\cite{KingmaB14} with a learning rate of $ 10^{-4} $ and weight decay of $ 10^{-6} $ and train for 240 epochs. The resulting precision-recall curve is shown in Figure~\ref{fig:dnn_prc}.
We also trained the classifier with several different random splits and with a split done in time (i.e., the most recent examples become the test set). We found no significant performance difference due to different seeds; however, splitting in time does cause the F1 score to drop to \( 0.78 \) from \( 0.88 \). We suspect that this is due to a notable difference in the anomaly percentage between the train and test sets when splitting in time: the train set is \( 21.2\% \) anomalous while the test set is only \( 12.5\% \) anomalous. Random splits of the dataset do not exhibit this same discrepancy. The practical problem of dealing with data drift in the RF station anomaly dataset, such as anomaly fraction, is deferred to a future work.

\clearpage

\bibliography{klysfault}

\end{document}